# Disentangling electronic and phononic contributions to high-temperature superconductivity in $X_2MH_6$ hydrides

Feng Zheng[1], Shiya Chen[2], Zhen Zhang[3], Renhai Wang[4], Feng Zhang[3,5], Zi-zhong Zhu[2], Cai-Zhuang Wang[3,5], Vladimir Antropov[3,5], Yang Sun[2, *], Kai-Ming Ho[3]

[1]*School of Science, Jimei University, Xiamen 361021, China*

[2]*Department of Physics, Xiamen University, Xiamen 361005, China*

[3]*Department of Physics and Astronomy, Iowa State University, Ames, Iowa 50011, USA*

[4]*School of Physics and Optoelectronic Engineering, Guangdong University of Technology, Guangzhou 510006, China*

[5]*Ames National Laboratory, U.S. Department of Energy, Ames, Iowa 50011, USA*

[*]Corresponding author: yangsun@xmu.edu.cn

## Abstract

Understanding the factors that control superconductivity is essential for discovering new superconducting materials using high-throughput elemental substitution. Focusing on the recently predicted ambient-pressure superconducting $X_2MH_6$ family, we disentangle the phononic and electronic contributions to $T_c$ to determine how isoelectronic substitution alters superconductivity. While substitution affects both phononic and electronic properties, the electronic contribution plays the dominant role in determining $T_c$ in the $X_2MH_6$ family. We show that the electronic contribution is affected by three key factors: the X–H bond distance, the electron localization function networking value of hydrogen, and the hydrogen-projected density of states at the Fermi level. A combined figure of merit derived from these parameters exhibits a robust correlation with $T_c$ across the family. We further show that pressure produces competing effects on superconductivity: it enhances the electronic contribution by shortening X–H bonds, but simultaneously weaken the phononic contribution by increasing phonon frequencies. The net pressure dependence of $T_c$ therefore results from the balance between these opposing tendencies. By disentangling and analyzing the electronic and phononic mechanisms, this work provides comprehensive insight into superconductivity in $X_2MH_6$ hydrides and offers practical guidance for designing new high-$T_c$ hydride superconductors.

# 1. Introduction

Hydrogen-rich compounds with high phonon frequencies and strong electron-phonon coupling are promising candidates for high-temperature superconductivity. Several binary hydrides, including $H_3S$[1, 2], $LaH_{10}$[3, 4], $YH_9$[5-7], $YH_6$[5] and $CaH_6$[8, 9], have been predicted and subsequently experimentally confirmed to exhibit very high $T_c$ ( > 200 K) at megabar pressure. However, the extremely high pressures required to stabilize superconductivity in these hydrides hinder their practical applications. Consequently, identifying hydride superconductors with high $T_c$ that are stable at lower or even ambient pressures has become a major research priority.

Compared with binary hydrides, ternary hydrides offer a much larger configurational space and thus greater potential for achieving higher $T_c$ or lowering the pressure required for high-$T_c$ superconductivity. A few ternary systems have been theoretically predicted or experimental synthesized to be high-$T_c$ superconductors at reduced pressures by introducing elemental substitution into binary hydrides[10-23]. Notably, high-throughput screening and machine-learning-assisted materials discovery have uncovered several new classes of ternary hydrides predicted to be promising high-$T_c$ candidates at ambient pressure, including the cubic $X_2MH_6$ (X is the group elements including alkali metals, alkaline earth metals and Aluminum; M is $3d$, $4d$ and $5d$ transition metal) family[24-28], $Mg_4Pt_3H_6$[29], $PdCuH_x$[30], hydride perovskites[28], hydride double perovskites[28], and other hydride structures[28]. These discoveries have expanded the structural database of hydride superconductors and are expected to stimulate experimental efforts[31] to realize high-$T_c$ hydride superconductors under ambient conditions.

Among the newly predicted ambient compounds, the $X_2MH_6$ systems have attracted particular attention due to their relatively high $T_c$ values (exceeding 50 K) with chemically abundant species, such as $Mg_2IrH_6$[24-26], $Li_2CuH_6$[26], $Li_2AgH_6$[32], $Li_2AuH_6$[26, 27], $Mg_2PtH_6$[25, 26], $Mg_2PdH_6$[25], $Mg_2RhH_6$[25, 26] and $Al_2MnH_6$[26]. A common feature of these superconducting compounds is their similar 19 or 20 valence electrons[28], suggesting that the valence-electron count is an important factor in determining superconductivity in $X_2MH_6$.

However, valence-electron count is a necessary but not sufficient condition for achieving high $T_c$ at ambient pressure. Isoelectronic substitutions were found to produce markedly different superconducting behaviors: for example, $Mg_2IrH_6$ is calculated to have $T_c$ = 117 K, whereas replacing Mg with Ca or Sr yields compounds with $T_c \sim$ 0 K[20]. Similar, $Li_2AuH_6$ has $T_c$ = 113 K, while $Na_2AuH_6$ and $K_2AuH_6$ show much lower $T_c$ values of about 53 K and 8 K, respectively[20]. These contrasts raise a fundamental question: what controls the divergent superconducting properties of isoelectronic $X_2MH_6$ compounds? Recent work by Wang et al.[33] analyzed the electronic structures of $Mg_2IrH_6$ and $Ca_2IrH_6$ and highlighted the critical role of the X-site ion. They argued that low-lying unoccupied Ca $d$ orbitals favor electron donation from the antibonding molecular $e_g^*$ orbitals of the $IrH_6^{4-}$ unit; because $IrH_6^{4-}$ motifs are critical to the superconducting mechanism, this back-donation into Ca $d$ states suppresses superconductivity. This argument was proposed to explain why $Mg_2IrH_6$ is superconducting while $Ca_2IrH_6$ is not and why high $T_c$ appear only when X-site ions do not have low-lying unoccupied $d$ orbitals, such as in $Mg_2IrH_6$, $Mg_2PtH_6$, $Mg_2RhH_6$, $Li_2CuH_6$, $Li_2AuH_6$ and $Al_2MnH_6$.

However, even when X-site ion lacks low-lying unoccupied $d$ orbitals as in both $Li_2AuH_6$ and $Na_2AuH_6$, their $T_c$ values differ by nearly 60 K (113 K vs 53 K)[20]. Moreover, we found that isoelectronic substitutions at M site can also produce large $T_c$ variations, such as $Mg_2IrH_6$ ($T_c$ = 117 K) versus $Mg_2CoH_6$ ($T_c$ = 36 K), or $Mg_2PtH_6$ ($T_c$ = 108 K) versus $Mg_2NiH_6$ ($T_c$ = 29 K) [20]. These contrasts highlight the complex interplay of chemical and electronic factors controlling superconductivity in $X_2MH_6$ and motivate a further analysis to uncover the mechanism that enable high $T_c$ in this family. To elucidate the superconductivity in these isoelectronic $X_2MH_6$ compounds, we separate the factors determining $T_c$ into phononic and electronic contributions and examine them independently. Both contributions are affected by elemental substitution, but the electronic contribution is more strongly influenced. The electronic behavior correlates closely with three tunable parameters — the X-H bond distance ($D_{X-H}$), the electron localization function (ELF) networking value $\phi$ of H, and the hydrogen-projected density of states at the Fermi level ($PDOS_H$) — which are modified by elemental substitution and thus control the electronic contribution to $T_c$.

We also find that pressure produces competing effect on the electronic and phononic contributions to superconductivity, and the balance of these effects ultimately determines different pressure dependence of $T_c$ in each compound.

## 2. Computational methods

Structure optimizations, density-of-states calculations, and ELF[34] analyses were performed using the projector-augmented wave (PAW) representations[35] with density functional theory implemented in the Vienna *ab initio* simulation package (VASP)[36, 37]. Exchange and correlation energy was treated within the spin-polarized generalized gradient approximation (GGA) and parameterized by Perdew-Burke-Ernzerhof (PBE) formula[38]. Wave functions were expanded in plane waves up to a kinetic energy cut-off of 550 eV. Brillouin-zone integrations were approximated using the special $k$-point sampling of the Monkhorst-Pack scheme[39] with a $k$-point mesh resolution of $2\pi \times 0.02 Å^{-1}$. Lattice vectors and atomic coordinated were fully relaxed until the force on each atom was less than $0.01\ eV \cdot Å^{-1}$. The networking value ($\phi$) of ELF was obtained by progressively lowing the ELF threshold until a continuous 3D network spanning the H atoms was identified, following the procedure of Belli *et al*[40].

The full Brillouin-zone electron-phonon coupling (EPC) calculation was performed with the Quantum ESPRESSO (QE) code[41, 42] based on the density-functional perturbation theory (DFPT)[43]. The pseudopotentials (stringent norm-conserving set) from the PSEUDODOJO project[44] for PBE functional were used. The kinetic energy cutoffs were 100 Ry for wave functions and 600 Ry for potentials. The convergence threshold for self-consistency was $1 \times 10^{-15}\ Ry$. The charge densities were calculated on a $k$ mesh of $12 \times 12 \times 12$ for ternary $X_2MH_6$ hydrides. The dynamical matrices were computed on a $q$ mesh of $4 \times 4 \times 4$ with smearing width of 0.02 Ry. Convergence tests shown in Fig. S1 suggest these parameters are sufficient for the EPC calculation. The calculations of superconducting $T_c$ are based on the Eliashberg spectral equation $\alpha^2 F(\omega)$ defined commonly now as

$$\alpha^2 F(\omega) = \frac{1}{2\pi N(E_f)} \sum_{qv} \frac{\gamma_{qv}}{\hbar \omega_{qv}} \delta(\omega - \omega_{q,v}), \tag{1}$$

where $N(E_f)$ is the density of states at the Fermi level, $\omega_{qv}$ denotes the phonon frequency of the mode $v$ with wave vector **q**. $\gamma_{qv}$ is the phonon linewidth defined as

$$\gamma_{qv} = \frac{2\pi \omega_{qv}}{\Omega_{BZ}} \sum_{ij} \int d^3k |g_{k,qv}^{ij}|^2 \delta(\epsilon_{q,i} - E_f) \delta(\epsilon_{k+q,j} - E_f), \tag{2}$$

where $\epsilon_{q,i}$ and $\epsilon_{k+q,j}$ are eigenvalues of Kohn-Sham orbitals at given bands and vectors. **q** and **k** are wave vectors, and $i$ and $j$ denote indices of energy bands. $g_{k,qv}^{ij}$ is the EPC matrix element, which describes the probability amplitude for the scattering of an electron with a transfer of crystal momentum **q**, determined by

$$g_{k,qv}^{ij} = \left(\frac{\hbar}{2M\omega_{qv}}\right)^{1/2} \left\langle \Psi_{i,k} \left| \frac{dV_{SCF}}{d\hat{\mu}_{qv}} \cdot \hat{e}_{qv} \right| \Psi_{i,k+q} \right\rangle, \tag{3}$$

where $M$ is the atomic mass, $\hat{e}_{qv}$ is the phonon eigenvector. $dV_{SCF}/d\hat{\mu}_{qv}$ measures the change of self-consistent potential induced by atomic displacement. $\Psi_{i,k}$ and $\Psi_{i,k+q}$ are Kohn-Sham orbitals. The EPC constant $\lambda$ can be determined through summation over the first Brillouin zone or integration of the spectral function in frequency space,

$$\lambda = \sum_{qv} \frac{\gamma_{qv}}{\pi \hbar N(E_F) \omega_{qv}^2} = 2 \int_0^\infty \frac{\alpha^2 F(\omega)}{\omega} d\omega. \tag{4}$$

The superconducting $T_c$ is determined with the analytical McMillan equation modified by the Allen-Dynes equation[45, 46],

$$T_c = \frac{f_1 f_2 \omega_{log}}{1.2} \exp\left[\frac{-1.04(1+\lambda)}{\lambda(1-0.62\mu^*)-\mu^*}\right], \tag{5}$$

where $f_1$ and $f_2$ are two separate correction factors, which are functions of $\lambda$, $\omega_{log}$, $\bar{\omega}_2$ and $\mu^*$. The effective screened Coulomb repulsion constant $\mu^*$ is 0.1 in our calculations. The $\omega_{log}$ is the logarithmic average frequency

$$\omega_{log} = \exp\left[\frac{2}{\lambda} \int \frac{d\omega}{\omega} \alpha^2 F(\omega) \log \omega\right], \tag{6}$$

The $\bar{\omega}_2$ is computed as

$$\bar{\omega}_2 = <\frac{2}{\lambda} \int d\omega \alpha^2 F(\omega)\omega >^{1/2} \qquad (7)$$

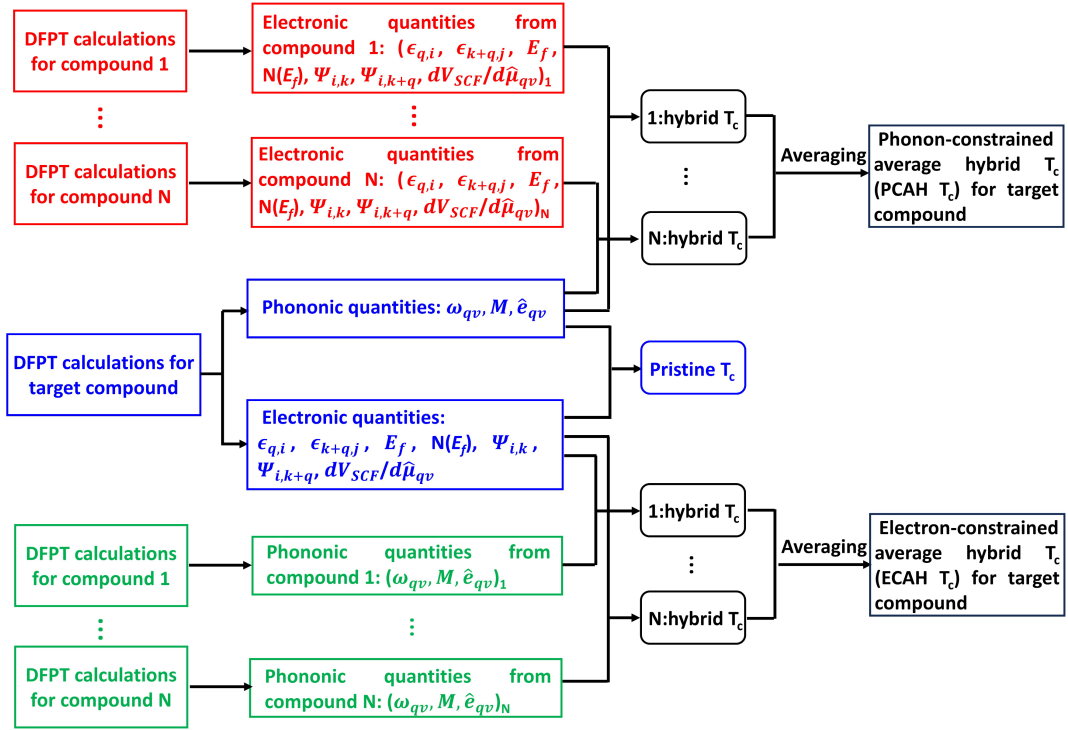

**Fig. 1.** Workflow for computing pristine and hybrid $T_c$s.

For a target compound, DFPT calculations yield a set of physical quantities that can be grouped into two components. The phononic component includes variables such as $\omega_{qv}$, M (atomic mass) and $\hat{e}_{qv}$. The electronic component comprises quantities such as $\epsilon_{q,i}$, $\epsilon_{k+q,j}$, $N(E_f)$, $E_f$, $\Psi_{i,k}$, $\Psi_{i,k+q}$ and $dV_{SCF}/d\hat{\mu}_{qv}$. Substituting these phononic and electronic variables into Eqs. (1)-(7) gives the superconducting $T_c$ for the compound. We refer to this value as the pristine $T_c$ because both phononic and electronic inputs are taken from the same material (highlighted in blue in Fig. 1).

To separate and quantify the individual effects of phonons and electrons on superconductivity, we define two averaged hybrid $T_c$ values: the phonon-constrained average hybrid $T_c$ (PCAH $T_c$) and electron-constrained average hybrid $T_c$ (ECAH $T_c$). For the PCAH $T_c$, we keep the target compound's phonon-related quantities fixed and combine them with the electronic quantities taken successively from N donor compounds (highlighted in red in Fig. 1), producing N hybrid $T_c$ values. We refer to these values as the hydride $T_c$s because the phononic and electronic inputs are taken from different materials. The mean of these hydride $T_c$s is the PCAH $T_c$. When different target compounds use the same set of N donor compounds for their electronic inputs, their PCAH $T_c$s are directly comparable. A higher PCAH $T_c$ therefore indicates a stronger phononic propensity for superconductivity for that compound. Conversely, for the ECAH $T_c$, we fix the electronic parameters of the target compound and pair them successively with the phonon parameters from N donor compounds (highlighted in green in Fig. 1). Averaging the resulting N hybrid $T_c$s yields the ECAH $T_c$. When different target compounds use the same set of N donor compounds for the phononic inputs, their ECAH $T_c$s are comparable. A larger ECAH $T_c$ thus indicates a stronger electronic contribution to superconductivity for that compound.

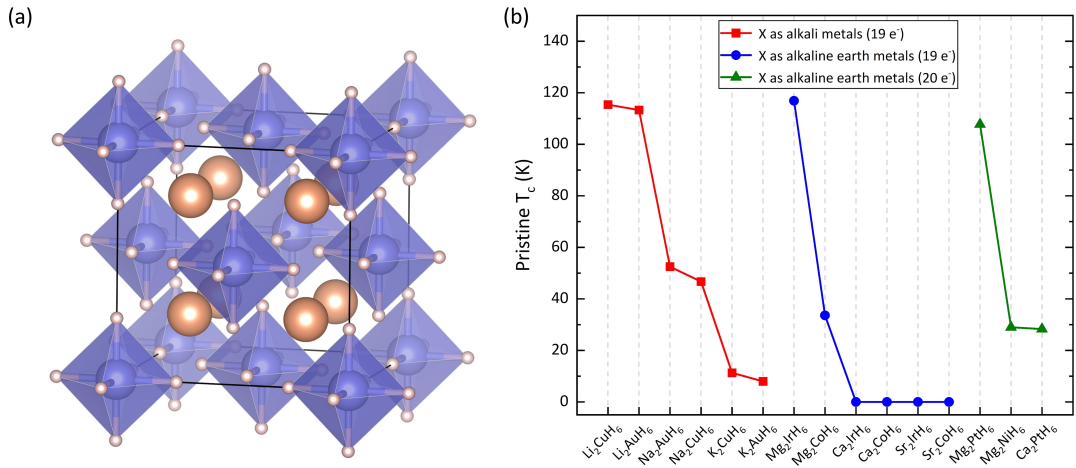

**Fig.2.** (a) Crystal structure of $X_2MH_6$; X, M, and H atoms are shown with orange, light blue, and white, respectively. (b) Calculated pristine $T_c$ for $X_2MH_6$ compounds. The red line denotes compounds with X as alkali metals (19 valance electrons); the blue and olive lines denote compounds with X as alkaline earth metals having 19 and 20 valance electrons, respectively.

## 3. Results and discussions

### 3.1 Variation in superconductivity of isoelectronic $X_2MH_6$ compounds

Fig.2 (a) shows the cubic crystal structure of $X_2MH_6$, featuring an $MH_6$ octahedron structural motif. By substituting the atoms at X and M sites, we generate a series of isoelectronic $X_2MH_6$ compounds. The stability of the $X_2MH_6$ family closely follows the valence-electron count. Taking $Mg_2MH_6$ as an example, Fig. S2 plots the lowest phonon frequency as a function of valence-electron count. At 18 ($Mg_2OsH_6$), 19 ($Mg_2IrH_6$), and 20 ($Mg_2PtH_6$) valence electrons, the softest mode lies at zero frequency, indicating that these phases are dynamically stable. For all other counts we examined, the lowest frequency is negative, indicating dynamically instability. Furthermore, as the electron count deviates further from 18–20, the lowest phonon frequency becomes increasingly negative, suggesting progressively reduced structural stability. To date, significant experimental efforts have been devoted to synthesizing one number of the $X_2MH_6$ family, $Mg_2IrH_6$. Although this phase is predicted to be synthesizable due to its small $E_d$ at ambient pressure[24-26], the synthesis route remains challenging. So far, two neighboring phases have been reported experimentally: $Mg_2IrH_5$ at low pressure [31] and cubic $Mg_2IrH_7$ above 40 GPa [47]. Because these share similar structures and bracket the $H_6$ composition, $Mg_2IrH_6$ appears as a natural intermediate and may be stabilized by non-equilibrium routes such as rapid pressure quenches, which may capture this stoichiometry before the system relaxes to ambient $Mg_2IrH_5$.

We consider 15 dynamically stable, isoelectronic $X_2MH_6$ compounds and compute the pristine $T_c$ for each. The $T_c$ values span a wide range (Fig. 2(b)). We divide them into three categories: (i) X = alkali metals with 19 valence electrons ($Li_2CuH_6$, $Li_2AuH_6$, $Na_2CuH_6$, $Na_2AuH_6$, $K_2CuH_6$, and $K_2AuH_6$); (ii) X = alkaline-earth metals with 19 valence electrons ($Mg_2CoH_6$, $Mg_2IrH_6$, $Ca_2CoH_6$, $Ca_2IrH_6$, $Sr_2CoH_6$, and $Sr_2IrH_6$); and (iii) X = alkaline-earth metals with 20 valence electrons ($Mg_2NiH_6$, $Mg_2PtH_6$, and $Ca_2PtH_6$). For the alkali-metal series (19 electrons), $T_c$ shows a stepwise trend: Li-containing compounds have the highest $T_c$, Na-containing ones are intermediate, and K-containing compounds show the lowest $T_c$. However, when X is fixed, substituting Cu by Au has little effect on $T_c$. In the alkaline-earth series (19 electrons), replacing Ir with Co for X = Mg causes a large drop in $T_c$ (from 117 K to 36 K), whereas Ca- and Sr-based compounds have $T_c$ values close to zero regardless of M. For the 20-electron series, a similar pattern appears: replacing Pt with Ni for X = Mg strongly reduces $T_c$, and changing the X ion from Mg to Ca also lowers $T_c$, although $Ca_2PtH_6$ retains a nonzero $T_c$ (unlike $Ca_2IrH_6$).

### 3.2 The effect of phononic properties on $T_c$ in the $X_2MH_6$ system

To understand why isoelectronic substitution in $X_2MH_6$ compounds can produce large changes in superconductivity (Fig. 2(b)), we first examine how phononic properties of these 15 compounds influence $T_c$. Following the workflow in Fig. 1, to ensure these PCAH $T_c$s comparability, we compute PCAH $T_c$s for each of the 15 target compounds using the same set of electronic donor compounds —i.e., the electronic inputs are drawn from these 15 compounds themselves.

As shown in Fig. 3, in the alkali-metal (19-electron) series the phonons of Li-containing compounds produce distributions shifted toward higher hybrid $T_c$ and yield larger PCAH $T_c$s than those of Na- and K-containing compounds, indicating that Li-based phonons are more favorable for superconductivity. However, the quantitative differences are modest: for $Li_2CuH_6$, $Na_2CuH_6$, and $K_2CuH_6$, the PCAH $T_c$s are ~54, ~40, and ~38 K, respectively, whereas their pristine $T_c$ values are 115, 47 and 11 K (Fig. 2(b)). Thus, although Li-containing phonons are somewhat more conducive to a high superconductivity, phononic effects alone cannot fully explain the large variation in pristine $T_c$ across the Li-, Na-, and K-containing compounds.

In the alkaline-earth (19-electron) series, $Mg_2IrH_6$ has the highest PCAH $T_c$, followed by $Mg_2CoH_6$, while Ca- and Sr-containing compounds have the lowest averages. The trend in PCAH $T_c$s among the Mg-, Ca- and Sr-containing compounds broadly mirrors that of the pristine $T_c$s, but the numerical discrepancies are substantial: for example, $Mg_2IrH_6$ and $Ca_2IrH_6$ have PCAH $T_c$s of ~35 K and ~22 K (Fig. 3), respectively, while their pristine $T_c$ values are 117 and 0 K (Fig. 2(b)).

Similar differences appear in the 20-electron alkaline-earth series, for instance, $Mg_2PtH_6$ and $Ca_2PtH_6$ have PCAH $T_c$s of ~57 K and ~36 K (Fig. 3) versus pristine $T_c$ values of 108 and 28 K (Fig. 2(b)). The discrepancies indicate that phonon properties alone cannot account for the observed pristine $T_c$ variations, so we next examine the role of electronic properties.

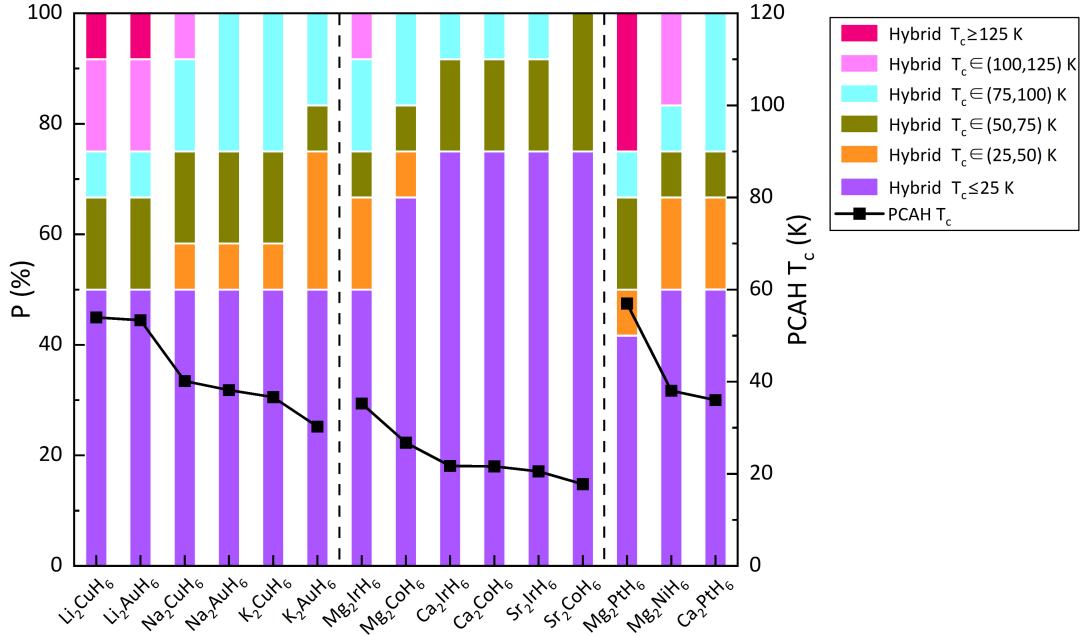

**Fig. 3.** Phononic favorability for superconductivity in $X_2MH_6$ compounds grouped into three categories delineated by dotted lines. The histogram shows the probability distributions of hybrid $T_c$ values with phononic quantities from target compounds. Black lines with square markers indicate the PCAH $T_c$s of the target compounds.

### 3.3 The effect of electronic properties on $T_c$ in the $X_2MH_6$ system

As with the phonon analysis, we assess how electronic properties of these 15 compounds influence $T_c$ by computing ECAH $T_c$s for each target compound following the workflow in Fig. 1. To ensure the comparability of these ECAH $T_c$s, we use the same set of phononic donor compounds for all calculations —i.e., the phononic inputs are drawn from these 15 compounds themselves.

As shown in Fig. 4, in the alkali-metal (19-electron) series the ECAH $T_c$s exhibit a clear stepwise trend: electronic parameters are most favorable for high hybrid $T_c$s in Li-containing compounds, followed by Na and then K. Compared with the PCAH $T_c$s (Fig. 3), the ECAH $T_c$s more closely reproduce the trend of the pristine $T_c$ values (Fig. 2(b)) arising from isoelectronic substitution of X in this series, indicating that changes in their pristine $T_c$ are primarily driven by variations in electronic properties. For a given X atom, Au generally provides slightly more favorable electronic properties for superconductivity than Cu (Fig. 4), whereas Cu yields slightly more favorable phononic properties (Fig. 3). Because superconductivity is determined by both electrons and phonons, their interplay can produce different outcomes: with X fixed, some compounds have marginally higher pristine $T_c$ values when M = Cu (e.g., $Li_2CuH_6$ vs. $Li_2AuH_6$), while in others favor M = Au (e.g., $Na_2AuH_6$ vs. $Na_2CuH_6$) (Fig. 2(b)).

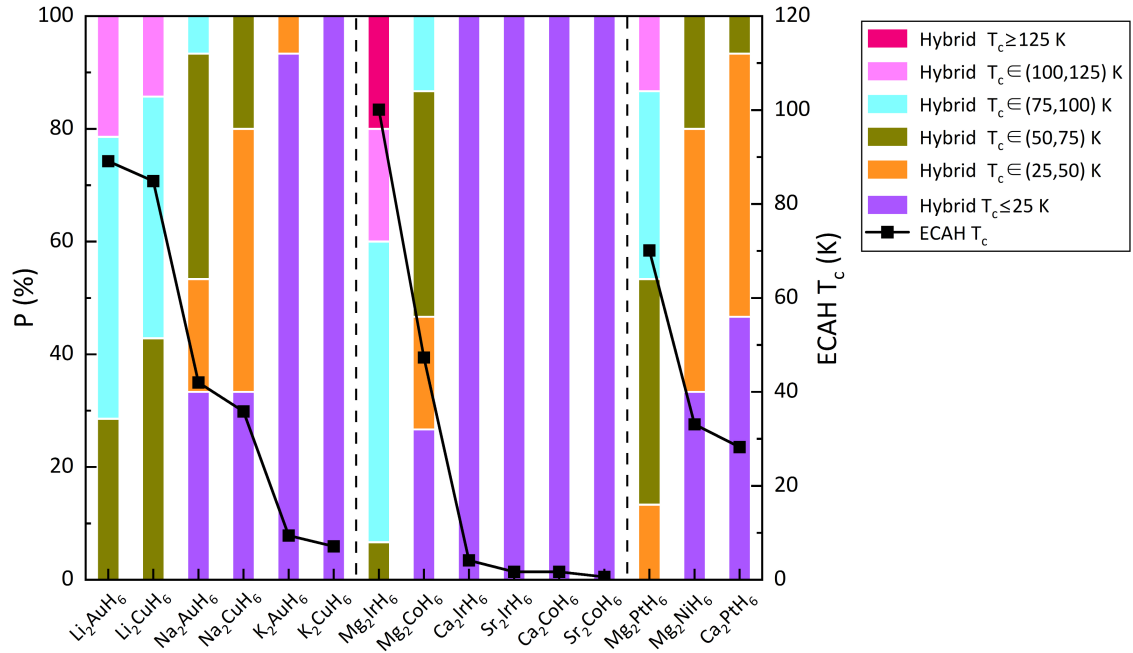

**Fig. 4.** Electronic favorability for superconductivity in $X_2MH_6$ compounds grouped into three categories delineated by dotted lines. The histogram shows the probability distributions of hybrid $T_c$ values with electronic quantities from target compounds. Black lines with square markers

indicate the ECAH $T_c$s of the target compounds.

In the alkaline-earth (19-electron) series, $Mg_2IrH_6$'s electronic structure strongly promotes superconductivity: when paired with any of the 15 phonon sets, all hybrid $T_c$s exceed 50 K, roughly 20% exceed 125 K, and the ECAH $T_c$ is ~100 K. Replacing Ir with Co reduces the ECAH $T_c$ to ~47 K, while substituting Mg with Ca or Sr drives the ECAH $T_c$s close to zero. Compared with PCAH $T_c$s (Fig. 3), the ECAH $T_c$s (Fig. 4) better reproduce both the trends and magnitudes of the pristine $T_c$ values from isoelectronic substitution (Fig. 2(b)), indicating that electronic properties dominate the observed changes in $T_c$ in this series. Thus, the suppression of superconductivity in $Ca_2CoH_6$, $Sr_2CoH_6$, $Ca_2IrH_6$ and $Sr_2IrH_6$ (Fig. 2(b)) can be mainly attributed to their unfavorable electronic properties.

A similar pattern holds in the alkaline-earth (20-electron) series. $Mg_2PtH_6$'s electronic properties are most conducive to high hybrid $T_c$ (ECAH $T_c \approx$ 70 K), whereas substituting Pt with Ni or replacing Mg by Ca produces electronic structures that are less favorable for superconductivity and lead to substantially lower ECAH $T_c$s of ~33 and ~28 K, respectively. Both the PCAH $T_c$s (Fig. 3) and ECAH $T_c$s (Fig. 4) capture the trend of pristine $T_c$ values (Fig. 2(b)), but the ECAH $T_c$s match the magnitudes more closely, confirming that electronic effects are the dominant factor.

### 3.4 Electronic origin of superconductivity variations upon substitution in $X_2MH_6$ compounds

From the above analysis, although both phononic and electronic changes can alter $T_c$ after isoelectronic substitution in $X_2MH_6$ compounds, the electronic contributions are generally the dominant factors. To elucidate the origin of these electronic contributions, we introduce an empirical parameter of electronic contributions (EC) to superconductivity, defined in Eq (8):

$$EC = PDOS_H \times \frac{\phi^2}{D_{X-H}^2} \quad (8)$$

where the PDOS$_H$ is the hydrogen-projected density of states at the Fermi level ($E_f$). A larger PDOS$_H$ means more electronic states at $E_f$ are available to participate in EPC, which enhances superconductivity. $\phi$ (networking value) is defined by Belli *et al.*[40] as the electron localization function (ELF) value of the largest isosurface that encloses H atoms (see in Fig. S3 for an example of the networking-value construction). A larger $\phi$ indicates a better-connected electronic network around H atoms that is highly sensitive to atomic displacements, which increases the electron-phonon matrix elements $g$; because these matrix elements enter the EPC strength quadratically, an increase in $\phi$ enhances the EPC constant roughly as $\phi^2$ and thus tends to raise electronic contributions to $T_c$. D$_{X-H}$ is the X-H bond distance. Shorter X-H bonds strengthen the nuclear potential felt by the electrons, so phonon-induced atomic displacements produce larger, steeper fluctuations in the local self-consistent potential ($dV_{SCF}/d\hat{\mu}_{qv}$). This derivative also enhances the electronic contribution to the EPC constant in a squared manner, so the electronic contribution to superconductivity is taken to be inversely proportional to D$_{X-H}^2$.

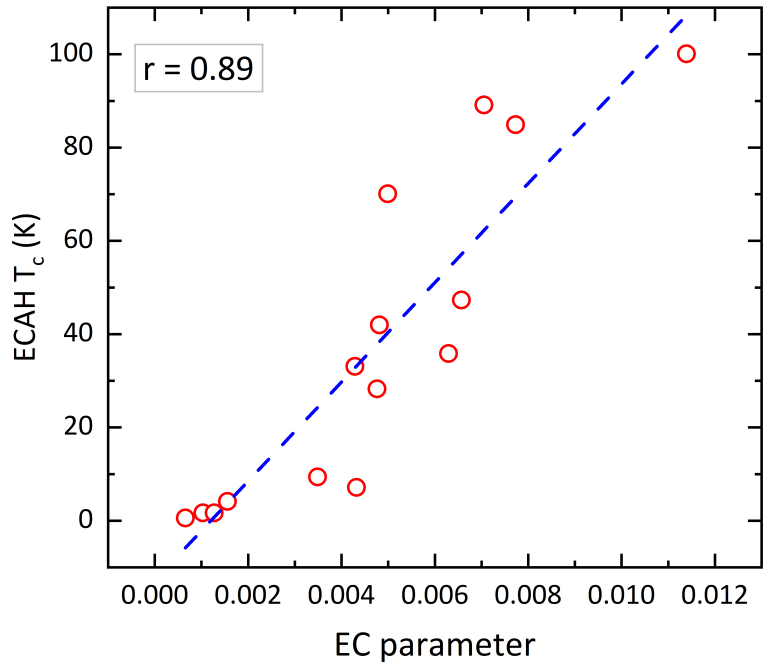

**Fig. 5.** Scatter plot showing the linear correlation between the EC parameter and the ECAH $T_c$s across X$_2$MH$_6$ compounds.

Although D_X-H, ϕ, and PDOS_H show no strong correlation with each other (Fig. S4-S6), the EC parameter displace a strong positive correlation with ECAH $T_c$s across the $X_2MH_6$ compounds (Pearson correlation coefficient $r = 0.89$) as shown in Fig. 5. This EC parameter therefore offers a quantitative metric for superconductivity from electronic contribution in the $X_2MH_6$ family. To further assess the robustness of the EC parameter, we applied the same EC definition to the recently reported transition metal hydrides superconductors $Li_3IrH_9$ [48] and its Na-substituted analogue $Na_3IrH_9$ at 25, 50 and 75 GPa. The EC parameter correlates strongly with their ECAH $T_c$s as well ($r = 0.98$; Fig. S7), supporting its use as a quantitative descriptor of superconductivity from electronic contribution in transition metal hydrides.

The variations in the electronic contribution to superconductivity across the $X_2MH_6$ family — and hence the observed differences in pristine $T_c$, since electronic properties largely govern pristine $T_c$ changes — can be traced to the three factors in Eq. (8) that change with elemental substitution. In particular, replacing the X-site atom with a large species increases the X-H distance ($D_{X-H}$) (Fig. S8), which lowers EC parameter (see Eq. (8)). This accounts for why Li-containing compounds have electronic properties more favorable for high pristine $T_c$ than their Na- and K-containing counterparts, and why Mg-containing compounds are more conductive to high pristine $T_c$ than Ca- and Sr-containing ones (Fig. 2(b)).

The unfavorable electronic contributions to superconductivity in $Ca_2CoH_6$, $Ca_2IrH_6$, $Sr_2CoH_6$, and $Sr_2IrH_6$ (Fig. 4), which produce pristine $T_c$ values near zero (Fig. 2(b)), reflect not only to their larger $D_{X-H}$ (Fig. S8(b)) but also to a reduced $PDOS_H$ at $E_f$ (Fig. S9) relative to $Mg_2CoH_6$ and $Mg_2IrH_6$. Substituting Ir with Pt in $Ca_2IrH_6$ leaves $D_{X-H}$ and $\phi$ essentially unchanged (Fig. S10(a)), but increases the $PDOS_H$ at $E_f$ in $Ca_2PtH_6$ (Fig. S10 (b)), yielding higher ECAH $T_c$ (Fig. 4). This explains why $Ca_2PtH_6$ is superconducting (pristine $T_c = 28$ K) while $Ca_2IrH_6$ is not, despite $Ca_2PtH_6$ having a smaller pristine $T_c$ than its Mg counterpart (Fig. 2(b)).

Replacing Co with Ir in $Mg_2CoH_6$ raises the $PDOS_H$ at $E_f$ from 0.216 to 0.286 st./eV/f.u. and increases the $\phi$ from 0.39 to 0.47 as shown in Fig S11. Although $Mg_2IrH_6$ has a slightly larger $D_{X-H}$ than $Mg_2CoH_6$ (2.35 vs 2.25Å), which is unfavorable for the electronic contribution to superconductivity, the combined figure of merit built from these three parameters nonetheless gives Mg$_2$IrH$_6$ a higher ECAH $T_c$ (~100 vs ~47 K; Fig. 4). Together with more favorable phononic properties (PCAH $T_c$ ~35 vs ~27 K Fig.3), this yields a much larger pristine $T_c$ for Mg$_2$IrH$_6$ than for Mg$_2$CoH$_6$ (117 vs 36 K; Fig. 2(b)).

In the X$_2$MH$_6$ system, isoelectronic substitution alters both phonon and electronic properties. With the structural framework largely unchanged, substitution primarily modifies bond lengths and lattice constants, which in turn shift the phonon dispersion; however, the impact on superconductivity is modest. In contrast, the electronic structure, responds more strongly: the Fermi-level density of states, the electron localization function (ELF), and the fluctuations of the local self-consistent potential under atomic displacements all vary more with substitution. These changes govern the evolution of the superconducting response. For X$_2$MH$_6$, superconductivity is therefore tuned primarily through the electronic channel.

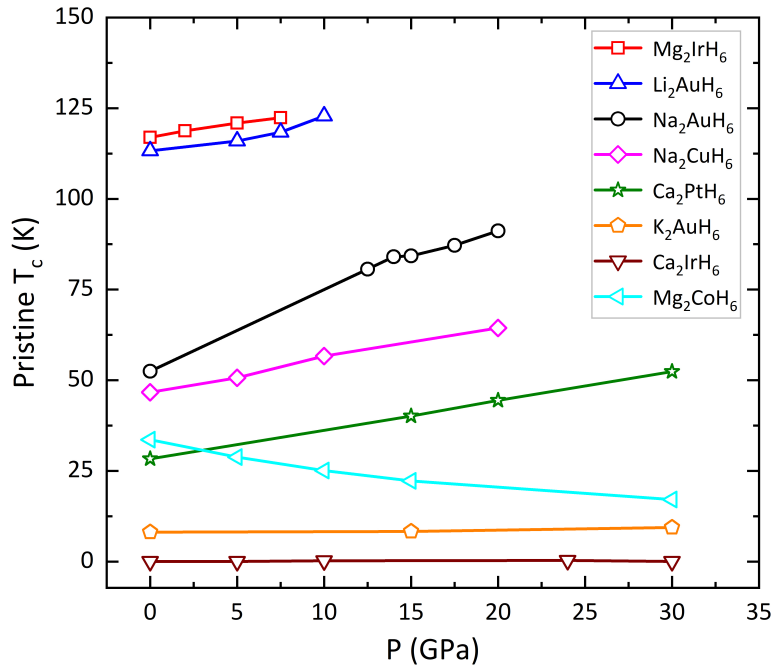

**Fig. 6.** Pressure-dependent pristine $T_c$ of X$_2$MH$_6$ ternary hydrides.

## 3.5 The effect of pressure on superconductivity of $X_2MH_6$ system

Because substituting the X-site atom with a larger species increases $D_{X-H}$ and thus reduces the electronic contribution to superconductivity (see Eq. (8)), it is natural to ask whether decreasing $D_{X-H}$ — for example by applying pressure — can improve the electronic properties and raise $T_c$. Indeed, as pressure increases, we find pristine $T_c$ rises for some compounds (e.g., $Li_2AuH_6$, $Na_2AuH_6$, $Na_2CuH_6$, $Mg_2IrH_6$, and $Ca_2PtH_6$), remains largely unchanged for others (e.g., $K_2AuH_6$ and $Ca_2IrH_6$) and decreases for $Mg_2CoH_6$ as shown in Fig.6. These contrasting pressure responds can be explained by the competing effects of pressure on both electronic and phononic parameters.

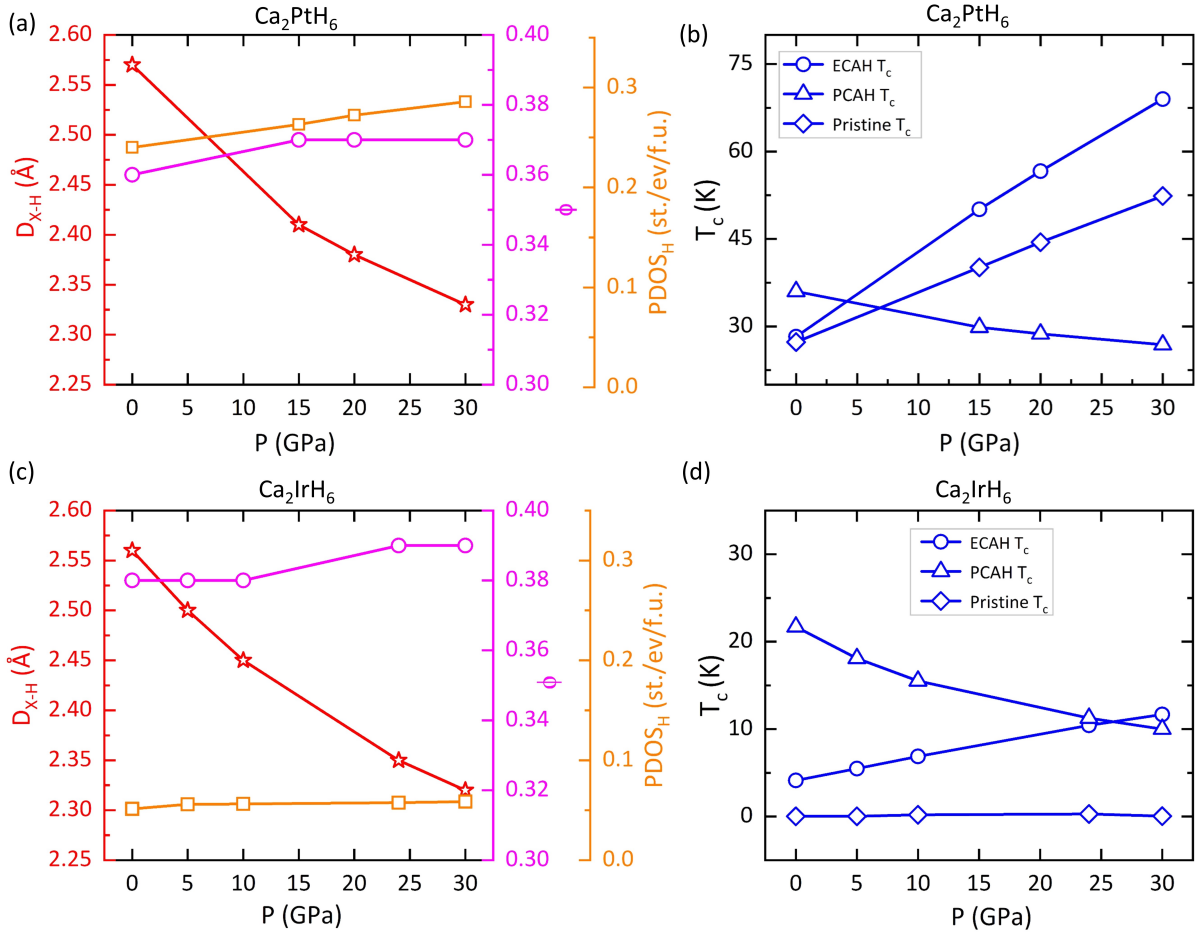

**Fig. 7.** X-H distance ($D_{X-H}$), the ELF networking value $\phi$, and the PDOS$_H$ at $E_f$ for (a) $Ca_2PtH_6$ and (c) $Ca_2IrH_6$ as functions of pressure. ECAH $T_c$s (computed with the same phononic inputs),

PCAH $T_c$s (computed with the same electronic inputs), and pristine $T_c$s for (b) $Ca_2PtH_6$ and (d) $Ca_2IrH_6$ as functions of pressure.

Taking $Ca_2PtH_6$ and $Ca_2IrH_6$ as examples. For $Ca_2PtH_6$ (Fig. 7(a)), increasing pressure only slightly increases the ELF networking value $\phi$ and the $PDOS_H$ at $E_f$, while progressively shortening $D_{X-H}$. According to Eq. (8), the reduced $D_{X-H}$ enhances the electronic contribution to superconductivity, and accordingly the ECAH $T_c$ of $Ca_2PtH_6$ rises with pressure (Fig. 7(b)). The enhanced electronic contribution can be reflected in the phonon linewidth γ: Fig. 8(a) and 8(b) show that γ of $Ca_2PtH_6$ at 30 GPa is substantially larger than at 0 GPa, particularly near the Γ point. This broadening results from the pressure-induced shortening of $D_{X-H}$, which enhances $dV_{SCF}/d\hat{\mu}_{qv}$, increases the electron-phonon matrix elements g, and thus widens γ.

However, increasing pressure raises the phonon frequencies of $Ca_2PtH_6$, especially in mid-to-low frequency range (Fig. 8(a) and 8(b)), which tends to reduce phononic contribution to $T_c$. This explains why the PCAH $T_c$ decreases with pressure (Fig.7(b)). Although pressure hardens the phonons, $\lambda$ nevertheless increases (from 0.80 at 0 GPa to 1.08 at 30 GPa (Fig. 8(a) and 8(b))) because the pressure-induced change in γ (the electronic term) is much larger than the change in ω (the phononic term) in Eq. (4). Thus, in $Ca_2PtH_6$ the electronic improvement under pressure outweighs the adverse phonon effects, so the overall pristine $T_c$ increases with pressure (Fig.6 and Fig. 7(b)).

Similarly, in $Ca_2IrH_6$, increasing pressure shortens $D_{X-H}$ while the ELF networking value $\phi$ and the $PDOS_H$ at $E_f$ increase only slightly (Fig. 7(c)). Because the $PDOS_H$ at $E_f$ is much lower than in $Ca_2PtH_6$, the comparable reduction in $D_{X-H}$ produces only a modest enhancement of electronic contribution to superconductivity, yielding a small rise in the ECAH $T_c$ for $Ca_2IrH_6$ (Fig. 7(d)). At the same time, pressure hardens the phonons in mid-to-low frequency range (Fig. 8(c) and 8(d)), which reduces the phononic contribution and causes the PCAH $T_c$ to fall (Fig. 7(d)). Because the electronic gain is largely offset by the phononic loss, the pristine $T_c$ of $Ca_2IrH_6$ remains essentially unchanged with pressure (Fig. 6 and Fig. 7(d)).

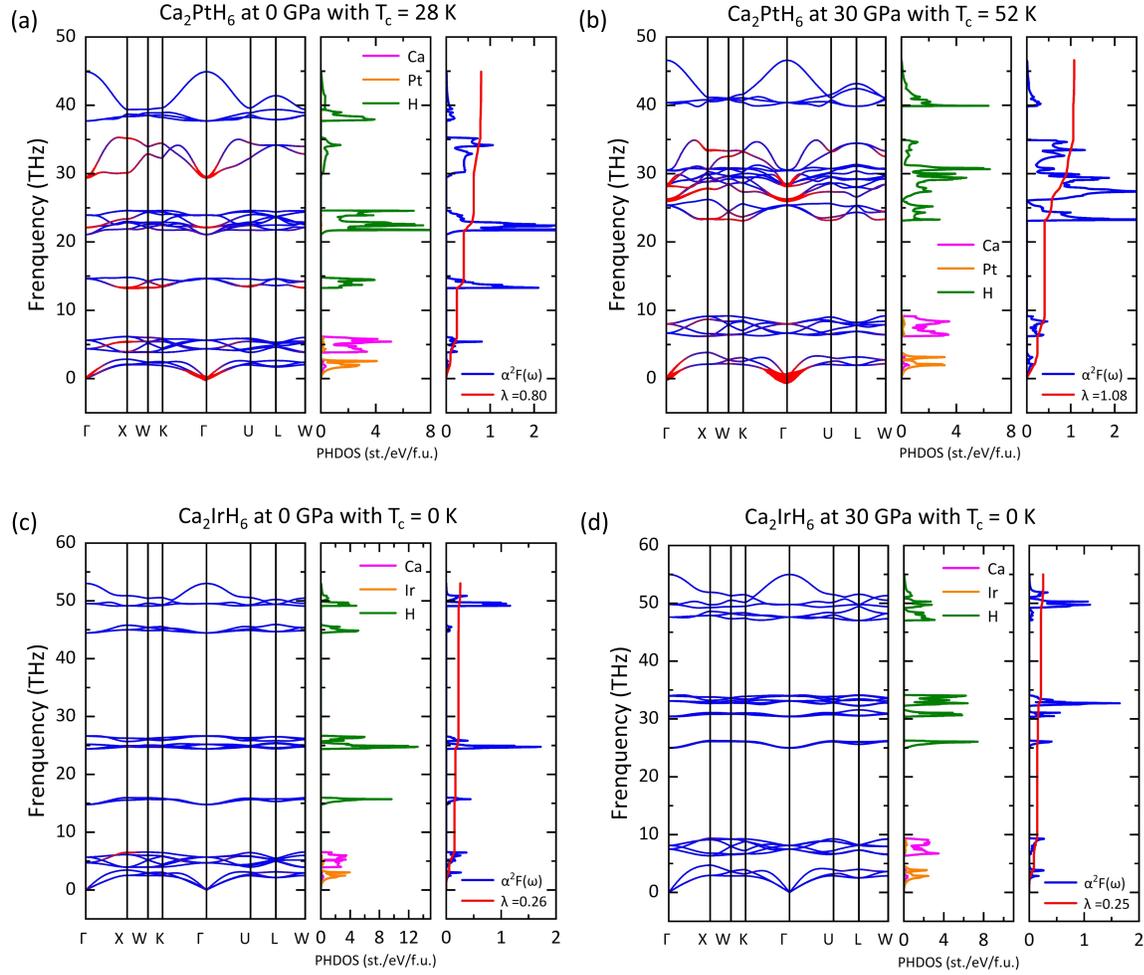

**Fig. 8.** The $\gamma_{qv}$-weighted (red dots) phonon spectrum, projected phonon density of states (PHDOS) and Eliashberg spectral function $\alpha^2 F(\omega)$ for $Ca_2PtH_6$ at (a) 0 GPa and (b) 30 GPa, and for $Ca_2IrH_6$ at (c) 0 GPa and (d) 30 GPa.

## 4. Conclusions

To explain why isoelectronic substitution produces markedly different superconducting behaviors across ternary $X_2MH_6$ hydrides, we separate the phononic and electronic effects of elemental substitution. Both channels affect $T_c$, but electronic contribution is dominant. Favorable electronic properties correlate strongly with a combined figure of merit composed of three parameters: a shorter X-H bond distance ($D_{X-H}$), a larger ELF networking value $\phi$ of H, and a higher H-projected $PDOS_H$ at $E_f$. Isoelectronic substitution modifies these three factors, thereby altering the electronic contribution and ultimately the pristine $T_c$. In particular, larger X atomic radii increase $D_{X-H}$ and thus weaken the electronic contribution, accounting for the trend of highest $T_c$ in Li-containing compounds, intermediate $T_c$ in Na-containing ones, and lowest $T_c$ in K-containing ones; Similar trends occur across the Mg-Ca-Sr series. When a large $D_{X-H}$ coincides with a reduced $PDOS_H$ at $E_f$, $T_c$ can be suppressed toward to zero, as in $Ca_2IrH_6$, $Ca_2CoH_6$, $Sr_2IrH_6$ and $Sr_2CoH_6$. We further find that pressure produces competing effect across $X_2MH_6$ compounds: it reduces $D_{X-H}$ and thus enhances the electronic contribution to superconductivity, but it also raises phonon frequencies and thereby weakens the phononic contribution. The balance among these competing factors determines the net impact of pressure on $T_c$ in each compound. Our study thus provides a clear picture of how elemental substitution and pressure influence superconductivity in $X_2MH_6$ family and offers guidance for designing new high-$T_c$ hydride superconductors.

## CRediT authorship contribution statement

**Feng Zheng:** Writing -review & editing, Writing – original draft, Validation, Methodology, Investigation, Project administration, Data curation, Funding acquisition. **Shiya Chen:** Investigation, Data curation. **Zhen Zhang:** Investigation, Formal analysis. **Renhai Wang:** Investigation, Data curation. **Feng Zhang:** Supervision, Formal analysis. **Zi-zhong Zhu:** Investigation, Formal analysis. **Cai-Zhuang Wang:** Supervision, Formal analysis. **Vladimir Antropov:** Investigation, Formal analysis. **Yang Sun:** Writing -review & editing, Supervision, Investigation, Formal analysis. **Kai-Ming Ho:** Supervision, Conceptualization.

## Data availability

Data will be made available on request.

## Acknowledgments

The work at Jimei University was supported by the National Natural Science Foundation of China (Grant No. 12404077), the Natural Science Foundation of Xiamen (Grant No. 3502Z202372015), the Natural Science Foundation of Fujian province (Grant No. 2024J01726) and the Research Foundation of Jimei University (Grant No. ZQ2023013). The work at Xiamen University was supported by the National Natural Science Foundation of China (Grant No. T2422016) and the Natural Science Foundation of Xiamen (Grant No. 3502Z202371007). The work at Guangdong University of Technology was supported by the National Natural Science Foundation of China (Grant No. 12504069). F.Z. and V.A. were supported by the U.S. Department of Energy, Office of Basic Energy Sciences, Division of Materials Science and Engineering. Ames National Laboratory is operated for the U.S. Department of Energy by Iowa State University under Contract No. DE-C02-07CH11358